\documentclass[]{aa}

\usepackage{aas_macros}
\usepackage{graphicx,txfonts,listings,ulem}
\usepackage{amsmath,amsfonts,amssymb,graphicx,color,times,bbm,psfrag,dsfont,slashed,bigints,bm,xcolor}
\usepackage{layouts}
\usepackage{subfigure}
\usepackage{multicol}
\usepackage[justification=justified]{caption}

\usepackage[colorlinks=true,citecolor=blue]{hyperref}

\begin{document} 
\title{Galactic diffuse gamma rays meet the PeV frontier}
%\title{The galactic gamma-ray diffuse emission meets the PeV frontier}
\subtitle{}
\author{P.~De La Torre Luque~\inst{1} %\email{mailto:pedro.delatorreluque@fysik.su.se}
\and
D.~Gaggero~\inst{4,5,6} 
%\email{mailto:daniele.gaggero@ific.uv.es}
\and
D.~Grasso~\inst{2,3}
%\email{mailto:dario.grasso@pi.infn.it}
\and
O.~Fornieri~\inst{7,8}
\and
K.~Egberts~\inst{9}
\and
C. Steppa~\inst{9}
\and
C.~Evoli~\inst{7,8}
}

\institute{
{Stockholm University and The Oskar Klein Centre for Cosmoparticle Physics, Alba Nova, 10691 Stockholm, Sweden}
\and
{INFN Sezione di Pisa, Polo Fibonacci, Largo B. Pontecorvo 3, 56127 Pisa, Italy}
\and
{Dipartimento di Fisica, Universit\`a di Pisa, Polo Fibonacci, Largo B. Pontecorvo 3}
\and
{Instituto de F\'isica Te\'orica UAM-CSIC, Campus de Cantoblanco, E-28049 Madrid, Spain}
\and
{Dipartimento di Fisica, Universit\`a di Torino,  via P. Giuria 1, I–10125 Torino, Italy}
\and
{Instituto de F\'isica Corpuscular, Universidad de Valencia and CSIC, Edificio Institutos de Investigac\'ion, Calle Catedr\'atico Jos\'e Beltr\'an 2, 46980 Paterna, Spain}
\and
{Gran Sasso Science Institute (GSSI), Viale Francesco Crispi 7, 67100 L’Aquila, Italy}
\and
{INFN-Laboratori Nazionali del Gran Sasso (LNGS), Via G.~Acitelli 22, 67100 Assergi (AQ), Italy}
\and
{Institut f\"ur Physik und Astronomie, Universit\"at Potsdam,  Karl-Liebknecht-Strasse 24/25, D 14476 Potsdam, Germany}
}

\date{}

\abstract
{The Tibet AS$\gamma$ and LHAASO collaborations recently reported the observation of a $\gamma$-ray diffuse emission with energy up to the PeV from the Galactic plane.}
{We discuss the relevance of non-uniform cosmic-ray transport scenarios and the implications of these results for cosmic-ray physics.}
{We use the {\tt DRAGON} and {\tt HERMES} codes to build high-resolution maps and spectral distributions of that emission for several representative models under the condition that they reproduce a wide set of local cosmic-ray data up to 100 PeV.}
{We show that the energy spectra measured by Tibet AS$\gamma$, LHAASO, ARGO-YBJ and Fermi-LAT in several regions of interest in the sky can all be consistently described in terms of the emission arising by the Galactic cosmic-ray ``sea''. We also show that all our models are compatible with IceTop $\gamma$-ray upper limits.} 
{Our results favor transport models characterized by spatial-dependent diffusion although some degeneracy remains between the choice of the transport scenario and that of the cosmic-ray spectral shape above 10 TeV. We discuss the role of forthcoming measurements in resolving that ambiguity.}


\maketitle
%-------------------------------------------------------------------

\section{Introduction}

The past few years witnessed remarkable observations in the field of multi-messenger astrophysics, especially in the cosmic-ray (CR) and $\gamma$-ray channels. The Tibet AS$\gamma$ experiment reported the observation of a diffuse $\gamma$-ray emission from the Galactic plane ($25^\circ < l < 100^\circ$ and $50^\circ < l < 200^\circ$, both for $\vert b \vert < 5^\circ$) with energy up to the PeV \citep{TibetASgamma:2021tpz}. This finding was confirmed by the LHAASO collaboration \citep{Zhao:2021dqj} for the innermost region. If not originated in currently-active unresolved sources (see {\it e.g.} \citealt{Casanova:2007cf} and the more recent \citealt{Vecchiotti:2021yzk} for an alternative interpretation), the presence of truly diffuse $\gamma$-rays at $\sim \mathrm{PeV}$ is likely due to $\sim \mathcal{O}(10) \, \mathrm{PeV}$ CRs injected by PeV accelerators that were active in the past, the so-called \textit{PeVatrons}. The ability to explore the \textit{knee} region $(E_{\mathrm{CR}} \sim \mathrm{few} \, \mathrm{PeV's})$ of the CR spectrum is of outstanding importance for our understanding of CR physics. Indeed, if we assume the conventional scenario of Supernova Remnants (SNRs) as the bulk of Galactic CRs (see \citealt{Gabici:2019jvz} for a recent review), it is a theoretical challenge to even achieve particle acceleration at the level of $\sim \mathcal{O}(100) \, \mathrm{TeV}$, and only for a short time in the evolution of the remnant~\citep{1983A&A...118..223L, 1983A&A...125..249L}. To overcome this problem, stellar clusters have recently come back as a viable explanation for such high-energy particle acceleration~\citep{1983SSRv...36..173C, 1:2021xpo}, although it is not clear up to what extent in the locally observed CRs. Therefore, whether the \textit{knee} in the CR spectrum is due to a change in the CR acceleration mechanism or to a transport effect is still matter of debate.

For what just described, the Tibet AS$\gamma$ and LHAASO measurements offer a new valuable handle to study the origin and the propagation of Galactic CRs, which complement direct measurements, that above $\sim \mathcal{O} (100) \, \mathrm{TeV}$ are performed with Extended Air Shower (EAS) experiments.
As witnessed by the large discrepancies in the energy spectra observed by different collaborations (see Fig. \ref{Fig1}), these measurements suffer from large systematic errors, mostly associated with modelling hadronic interaction within the Earth atmosphere. Another valuable advantage of measuring the $E_{\gamma} \geq \mathrm{TeV}$ $\gamma$-ray diffuse emission of the Galaxy is that it may allow to test whether features in the CR spectra --- such as the hardening at $\sim 300\,{\rm GeV/n}$ found by CREAM \citep{Ahn:2010gv}, PAMELA \citep{PAMELA:2011mvy}, AMS \citep{AMS:2015tnn,AMS:2015azc}  and the softening at $\sim 10\,{\rm TeV/n}$ measured by DAMPE \citep{DAMPE:2019gys} and CREAM \citep{Yoon:2017qjx} --- are due to local sources or if they are representative of the whole Galaxy, namely due to transport effects~\citep{Blasi:2012yr}.
Moreover, the detection of the diffuse emission at low Galactic longitudes gives crucial information on how CR propagation behaves in the inner regions of the Galaxy, where the conditions for CR transport are expected to be different than in the average disc, noticeably leading to a radial (Galactocentric) dependence for the CR spectra.
It should be noted that, while in the last years a huge experimental effort allowed to strongly reduce the uncertainties on the CR transport parameters within a few kpc's horizon around the Solar System, very little is still known about CR propagation beyond that distance. With this regard, the study of weakly interacting messengers such as $\gamma$-rays and neutrinos originated in the scattering of CRs with the InterStellar Medium (ISM) and the IS Radiation Field (ISRF) offers a valuable probe of CR transport in those distant regions.

In \citet{Gaggero:2014xla,Cerri2017jcap} the authors argue that the hardening of the $\gamma$-ray diffuse emission found by Fermi-LAT above 10 GeV \citep{Fermi-LAT:2012edv,Fermi-LAT:2016zaq} -- see also \citet{Yang2016prd} -- can be explained in terms of a radially dependent CR spectrum.
Based on that interpretation, in \citet{Gaggero:2015xza} the authors predict a diffuse flux that is much larger than what is expected by conventional transport models at low Galactic longitudes and very high energies $E \gg 1$ TeV. They show that this also allows to consistently reproduce Fermi-LAT and Milagro \citep{Abdo:2008if} measurements in the inner Galaxy,
as well as \citep{Gaggero:2017jts} the H.E.S.S. observed diffuse-emission in the Central Molecular Zone \citep{HESS:2016pst} (although the CR population in that small region may be affected by one or more PeVatrons close to the Galactic Center (GC)).

Recently, a phenomenological model inspired by that scenario --- where the energy and spatial dependence of the CR spectrum are {\it non-factorized} --- has been studied~\citep{Lipari:2018gzn} (LV non-factorized model hereafter) and used by Tibet AS$\gamma$ to interpret their results~\citep{TibetASgamma:2021tpz}.
For reference, the authors also considered a conventional (LV {\it factorized}) model with a radially independent CR spectrum.
%which, however, was found to be disfavored by the observations.
LV models also account for $\gamma$-ray absorption -- which is significant above 100 TeV -- due to pair production by scattering onto the CMB and 
%\melo{the IR galactic (come includono l'IR?)} 
the dust emitted infra-red photon background ~\citep{Vernetto2016prd}. 
Both LV models are built to reproduce AMS-02 and CREAM proton and Helium data below 100 TeV.
As shown in~\citet{TibetASgamma:2021tpz} the LV non-factorized (spatial dependent) model seems to be in better agreement with Tibet AS$\gamma$ results than the factorized (conventional) one. Preliminary LHAASO data \citep{Zhao:2021dqj}, however, are slightly lower than Tibet in some energy bins. Most importantly, they have significantly smaller statistical errors and extend down to 10 TeV where they are less scattered and the uncertainties on the CR spectral elemental shapes are smaller. As we will show this may have relevant implications for their interpretation.

In this work we model the $\gamma$-ray diffuse emission of the Galaxy from 10 GeV up to the PeV trying to assess the effect of the uncertainties due to the choice of the CR spectra, hence on the choice of the transport scenario.
%the main (light) components of the CR population above 10 TeV. 
%
Differently from what has been done in~\citet{Lipari:2018gzn}, we make extensive use of Fermi-LAT results -- showing their relevance to constrain CR transport properties -- and use physically motivated models.
In fact, CR transport is modelled with the {\tt DRAGON2} numerical code \citep{Evoli2017jcap,Evoli2018jcap}
in two transport scenarios: either assuming a spatially independent or dependent diffusion coefficient. 
Additionally, we use the newly released {\tt HERMES} code \citep{Dundovic:2021ryb} to integrate along the line of sight the CR spatial and energy distributions obtained with {\tt DRAGON2} using detailed IS gas emission maps, updated ISRF models and $\gamma$-ray production cross-sections, to get high resolution sky maps of the diffuse emission at all relevant energies, which allow to provide more accurate and robust results than with the analytical gas models used in \citep{Lipari:2018gzn}. 
Moreover we account for a wider set of CR data above 10 TeV (see, however, \citealt{Vernetto:2021tgp}) and, in order to cope with the large discrepancies among different data sets (see also \citealt{Koldobskiy:2021cxt}), we consider two set-ups for the CR injection spectra. 
We will then compare our predictions with Tibet AS$\gamma$ and LHAASO results as well as with ARGO-YBJ~\citep{ARGO-YBJ:2015cpa} at lower energy, and with IceCube~\citep{IceCube:2019scr} and CASA-MIA~\citep{Borione:1997fy} upper limits between 100 TeV and 2 PeV in regions closer to the GC, besides with Fermi-LAT data in several Galactic Plane (GP) quadrants. 

We will show that Tibet AS$\gamma$ and, if confirmed, LHAASO data in combination with Fermi-LAT 
%-- especially the bins between 10 and 100 TeV -- 
%, hence the most \melo{\sout{reliable ones} che vuol dire? intendi error bar piccoli?} 
favor a spatially dependent transport scenario.
%We will also discuss the perspectives of SWGO \citep{Albert:2019afb} to strengthen (or to reject) this conclusion.
%--------------------------------------------------------------------

\section{Relevant data}

The high-precision measurements provided by recent cosmic-ray missions such as AMS-02~\citep{Aguilar:2015ooa, Aguilar:2014mma, aguilar2015precision} and DAMPE~\citep{Ambrosi:2017wek, DAMPE_protons} together with the increasing quality and quantity of gamma-ray data, thanks to the Fermi-LAT~\citep{Ackermann_2012, Ackermann:2014usa}, HAWC~\citep{Albert:2020fua, Abeysekara:2018bfb} and H.E.S.S.~\citep{2008PhRvL.101z1104A, HESS_Coll2018} experiments, require improved models on particle acceleration and transport for their interpretation~\citep{Serpico:2018lkb}.

\subsection{Local cosmic-ray data}\label{sec:CRdata}

%{\it Gamma rays are mainly produced via the decay of the neutral pions produced in the hadronic showers initiated by the collision of a cosmic-ray (CR) proton or helium against the interstellar atomic $({\rm HI})$ or molecular $({\rm H}_2)$ gas. While at high Galactic latitudes leptonic processes are important --- due to a less dense medium and high magnetic field ---, the process $\pi^0 \rightarrow \gamma \gamma$ dominates the $\gamma$-ray sky $(E_{\gamma} > 1 \, \mathrm{GeV})$ on the Galactic plane in the inner as well as the outer regions of the Galaxy. Therefore, in order to achieve an accurate description of the $\gamma$-ray production, we study the hadronic, primary-CR population. The result of this procedure is shown in Figure \ref{Fig1}.}{\textcolor{red}{\it This paragraph may be removed since it has no connection with data !}}

In the low-to-intermediate energy range $(10 \, \mathrm{GeV} \lesssim E_{\mathrm{CR}} \lesssim 10 \, \mathrm{TeV})$, CR data are collected by space-born or balloon experiments, that are less affected by systematic uncertainties. 
In this energy band we consider the results from AMS-02~\citep{Aguilar:2015ooa, aguilar2015precision}, DAMPE~\citep{DAMPE_protons, Alemanno:2021gpb}, CALET~\citep{CALET2019}, ATIC-2~\citep{ATIC2009}, CREAM-III~\citep{CREAM2017} and NUCLEON~\citep{NUCLEON2018}. 

As shown in Fig.~\ref{Fig1}, measurements are in good agreement with each other for both protons and Helium nuclei, and show a power-law structure up to $\sim \mathcal{O}(100 - 500) \, \mathrm{GeV}$, where a spectral \textit{hardening} is reported by all the experiments, typically interpreted as a consequence of a new phenomenon occurring in particle transport rather than in the acceleration  (see \citealt{Gabici:2019jvz} and references therein). 
Another structure that is clearly visible in the proton and helium spectra is a pronounced \textit{softening} reported by DAMPE at a rigidity of $\sim \mathcal{O}(10) \, \mathrm{TV}$~\citep{DAMPE_protons, Alemanno:2021gpb}, confirming previous claims by CREAM-III~\citep{CREAM2017} and  NUCLEON~\citep{NUCLEON2018}. 

It is important to notice that, since the highest-precision measurements come from experiments that are not ground based -- for which it is challenging to achieve energies above $\sim 10 \, \mathrm{TeV}$ with significant statistics --, the highest energies currently reached do not allow to discriminate whether the softening is a (i) local feature or (ii) a global structure.  In the first scenario (i), the particle spectra in the Galaxy would follow its trend after the hardening at a few hundreds of GeV and proceed up to the \textit{knee} at $\sim 5 \, \mathrm{PeV}$, where it undergoes a cut-off in the form of a sharp softening --- the ``bump" is then typically interpreted as a signature of a nearby source contributing to the primary CRs~\citep{Fornieri:2020xai}. Throughout the paper, we refer to this setting as the \textit{Max} setup. As for the second case (ii), we would expect a softening above $E_{\mathrm{CR}} \sim \mathcal{O}(10) \, \mathrm{TeV}$ and a cutoff as a pronounced break at the \textit{knee} (at $\sim$ a few PeV). We refer to this other setting as the \textit{Min} setup. 
%Higher-energy data points in the forthcoming years will allow to disentangle these two pictures.

For what concerns the highest-energy band $(1 \, \mathrm{PeV} \lesssim E_{\gamma} \lesssim 1 \, \mathrm{EeV})$, such energies can be achieved at the moment only by ground-based experiments measuring \textit{extensive air showers} (EAS), which imply much larger uncertainties and strong dependence on models of the hadronization occurring in the Earth's atmosphere, specifically on their Monte Carlo-based simulations  (see \citealt{HadronicUncertainties2006, PARSONS2011832} for a discussion on the model-dependent uncertainties of the hadronic showers). In particular, we are considering data points from KASCADE~\citep{KASCADE2005}, KASCADE-Grande~\citep{KASCADEgrande2013} or, in alternative, from IceTop~\citep{IceTop2019}, being the two datasets incompatible with each other.

In what follows, we compute the $\gamma$-ray emission in the Galaxy for both the \textit{Min} and the \textit{Max} spectral setups. As shown in Fig.~\ref{Fig1}, our procedure is addressed to bracket the uncertainties due to the poor knowledge of the CR spectrum in the highest-energy range.

As an additional comment, we do not account for CR species heavier than Helium since, under reasonable conditions, their contribution to the gamma-ray source term is subdominant (less than 10\%) with respect to the expected flux due to proton + Helium production rate (see {\it e.g.} \citealt{Breuhaus:2022epo} for a recent analysis).

\subsection{Fermi-LAT data}\label{sec:Fermi_data}
The LAT is a pair-conversion telescope that provides gamma-ray data from August 2008 in a range of energies ranging from $\sim20$~MeV to a few hundreds GeV \citep{Atwood2009apj}.
In this work, we make use of $\sim 149$ months of data (from 2008-8-2008 to 2020-12-31), selecting CLEAN events from the PASS8 data. We select events from good quality time intervals and remove the intervals when the LAT was subtended at rocking angles $\theta_r < 52\deg$ ((DATA\_QUAL>0) \&\& (LAT\_CONFIG==1) \&\& ABS(ROCK\_ANGLE<52)), and in addition apply the zenith-angle cut $\theta_z < 100\deg$, in order to avoid contamination with the Earth \textit{albedo}. We consider front- and back-converted events in order to minimize statistical errors. We employ the P8R3\_CLEAN\_V3 version of the instrument response functions. The extraction of Fermi data and calculation of exposure maps is performed using the most up-to-date version of the ScienceTools\footnote{https://fermi.gsfc.nasa.gov/ssc/data/analysis/software/; https://github.com/fermi-lat/Fermitools-conda/wiki/Installation-Instructions} (2.0.8; released on 01/20/2021). 
%{\color{red} Reference to the APPENDIX }
In this work, our $\gamma$-ray model is optimized to reproduce Fermi-LAT data in different windows of galactic longitude in the energy range between $\sim10$ to $\sim300$~GeV, under the hypothesis of a spatially-dependent spectral index of the diffusion coefficient (see the details in section~\ref{sec:CRtransp}), considering the recent Fermi 4FGL-DR2 catalog for accounting for the source's $\gamma$-ray emission.

%We also make use of the Isotropic spectral template provided by the Fermi-LAT collaboration (iso\_P8R3\_CLEAN\_V3\_v1)\footnote{https://fermi.gsfc.nasa.gov/ssc/data/access/lat/BackgroundModels.html} to account for extragalactic isotropic emission.

\subsection{Very High Energy gamma-ray data}\label{sec:gamma_data}
A dedicated section is necessary for the very-high energy band, namely for the photons detected with energy $E_{\gamma} \geq 100 \, \mathrm{TeV}$, that can be probed via the deconvolution of the EAS's at ground-based detectors. In particular, we want to comment on the observations of PeV $\gamma$-rays by the LHAASO~\citep{2021Natur.594...33C} and Tibet AS$\gamma$~\citep{TibetASgamma:2021tpz} Collaborations. LHAASO reports the detection of $\sim 530$ photons above $E_{\gamma} = 100 \, \mathrm{TeV}$ at $\sigma \geq 7$ significance from 12 regions with overlapping known sources. Among these, the source J2032+4102 --- associated with the stellar cluster Cygnus-OB2, considered a probable PeV-acceleration site~\citep{Aharonian:2018oau} --- deserves special attention, as it is where the highest-energy photon $(E_{\gamma} = 1.42 \pm 0.13 \, \mathrm{PeV})$ is originated. Alongside, Tibet observes 10 events in the energy bin $398 \leq E_\gamma(\mathrm{TeV}) \leq 1000$, 4 of which originate around the OB2 cluster. However, in order to remove the cluster contribution to the total detected photons, in Tibet they use a Gaussian profile with width $\sigma_{OB} = 0.5^\circ$ --- this is consistent with the angular extension exploited by LHAASO~\citep{2021Natur.594...33C} ---, motivated by the profiles $\sigma_{\mathrm{HAWC}} \leq 0.3^\circ$ used by the HAWC Collaboration for sources above $E_{\gamma} = 56 \, \mathrm{TeV}$~\citep{HAWC:2019tcx}. On the other hand, for the specific case of the Cygnus \textit{cocoon}, HAWC reports an extension of $\sigma_{\mathrm{OB, \, HAWC}} = 2.1^\circ$~\citep{Abeysekara:2021yum}, which implies that most of the diffuse emission from the Cygnus region could actually be coming from the single source. This information has been used in \citet{Liu:2021lxk} to interpret the neutrino counterpart expected from the region, that would otherwise be in tension with the IceCube observations on the GP~\citep{IceCube:2017trr}. 

As a result, it is possible that the highest energy $\gamma$-ray point in the Tibet observations actually should show a flux lower than what is currently estimated. Future analyses will certainly help to clarify the picture.

\begin{figure*}[h!]
   \centering
   \includegraphics[width=0.45\linewidth]{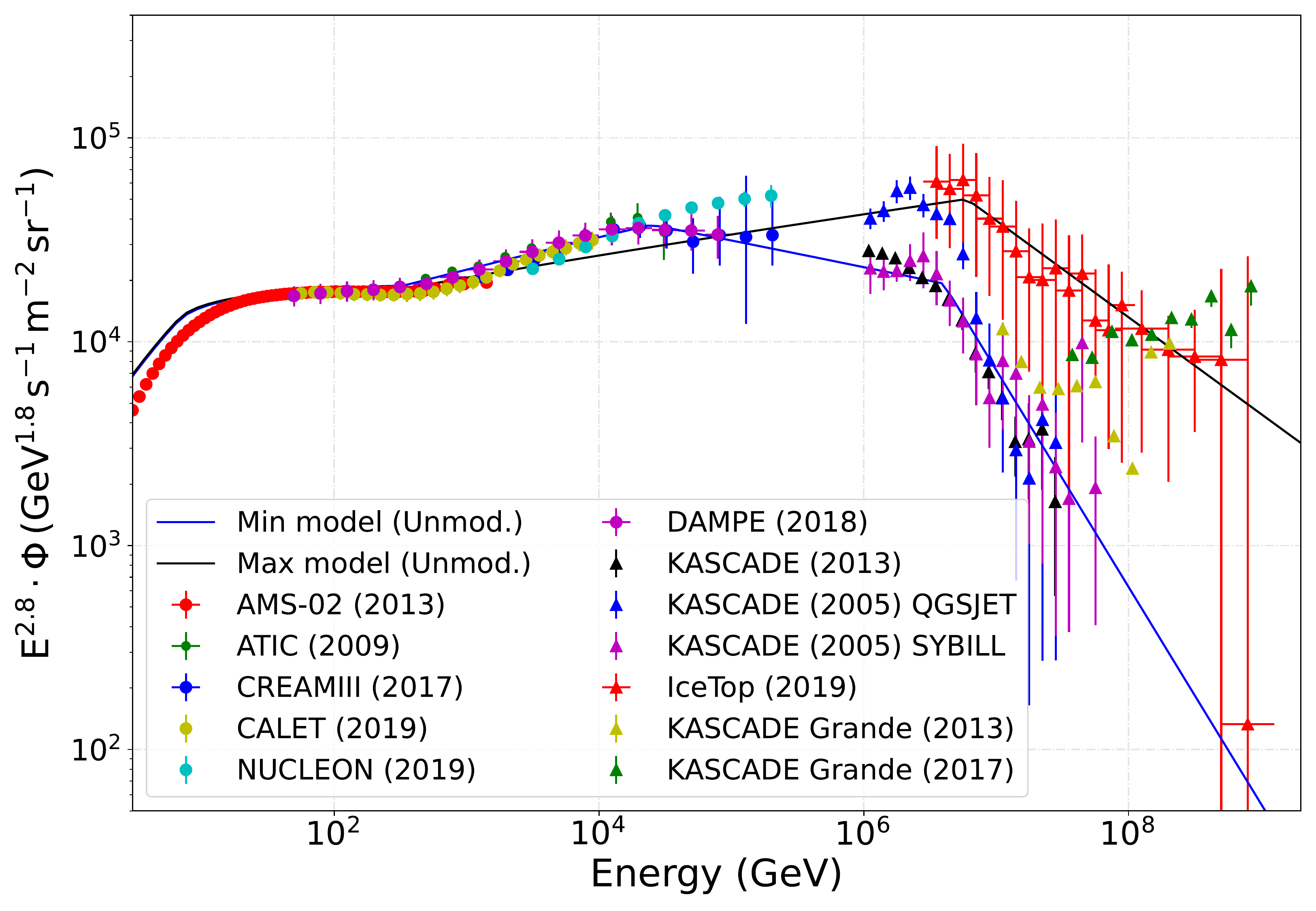}
   \includegraphics[width=0.45\linewidth]{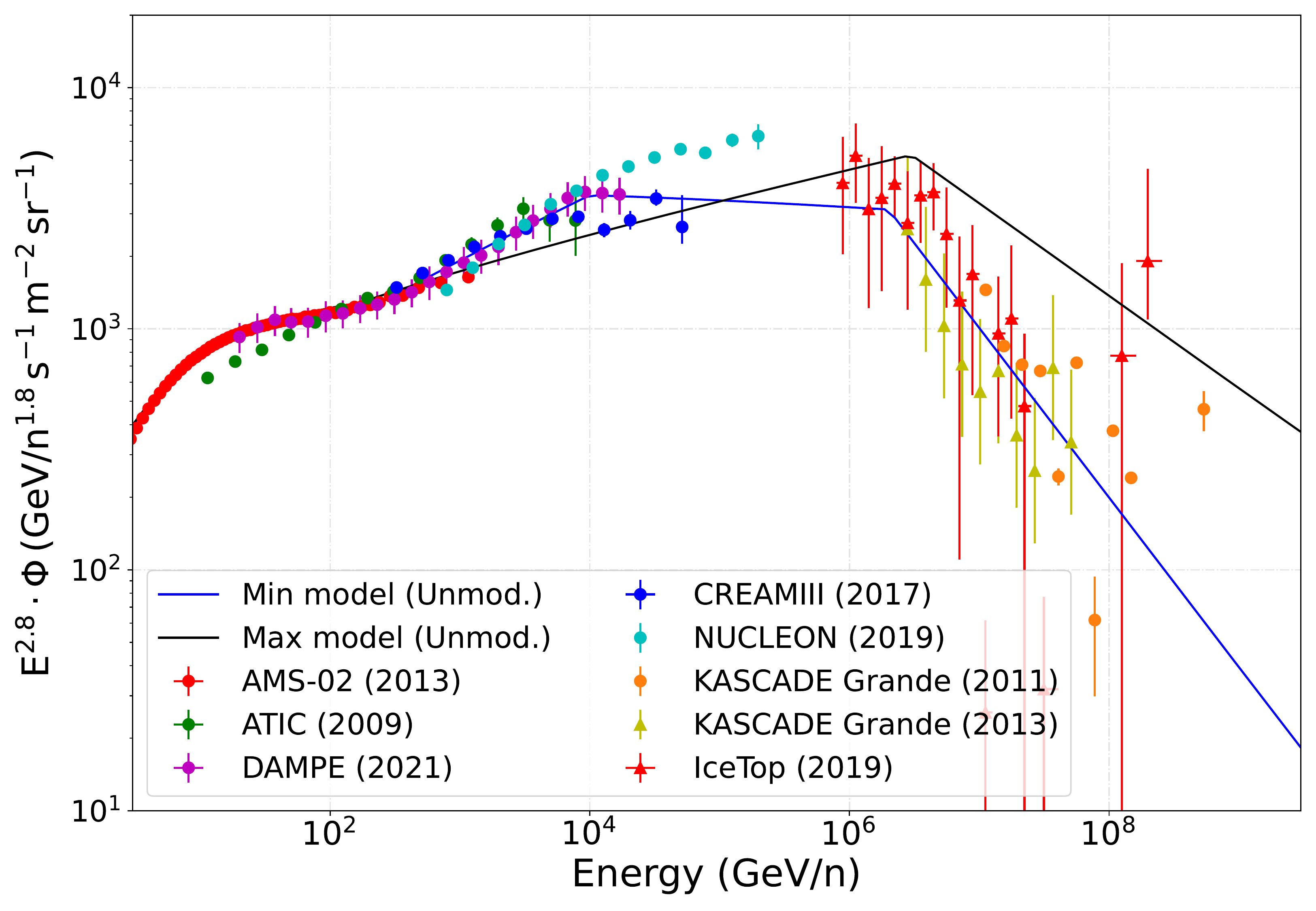}
   \caption{\small{The proton (left panel) and Helium (right panel) local spectra computed for the $\gamma$-optimized scenario are plotted against a representative set of data. For each species the spectra as predicted using the Max and Min source spectrum set-ups are shown. 
   We do not show here the corresponding lines computed for the Base scenario since they are almost coincident with those reported here above 10 GeV/n (at the Solar System position).
   %These models allow us to roughly bracket the experimental uncertainties around the PeV using two different interpretations of the spectra measured below this energy.
   }}
\label{Fig1}%
\end{figure*}

\section{The models}\label{sec:models}

The diffuse $\gamma$-ray sky is the superposition of a number of different components. They include
\begin{itemize}
    \item {\it the isotropic background}: This component captures the contribution from extra-galactic diffuse emission, unresolved extra-galactic sources, and residual (mis-classified) cosmic-ray emission.   %this is an almost isotropic emission due to the superposition of a large number of extra-galactic sources including AGN's, normal and star-burst galaxies, galaxy clusters. Fermi-LAT found this emission to extend up to few hundred GeV \citep{Fermi-LAT:2014ryh}.
    We adopt here a reference isotropic spectral template provided by the Fermi-LAT collaboration (iso\_P8R3\_CLEAN\_V3\_v1)\footnote{https://fermi.gsfc.nasa.gov/ssc/data/access/lat/BackgroundModels.html}.
    %to account for extragalactic isotropic emission.
    \item {\it the unresolved point-like, or extended, source emission}:  The contribution of unresolved sources is derived from population synthesis, using the model for the Galactic gamma-ray population following a four-arm spiral distribution described in \citet{SteppaEgberts2020}. This source model for fluxes at 1~TeV is translated to the energy range of the Fermi-LAT data using a spectral index of -2.4 for the TeV energy range as the mean spectral index of the H.E.S.S. Galactic source catalog~\citep{HESS_Coll2018} and for lower energies a spectral index of -2.27 as mean of the Fermi-LAT 4FGL~\citep{Fermi4FGL} with a transition regime between 100~GeV and 1~TeV. Source populations are simulated and their sources classified as detectable or unresolved according to the Fermi-LAT sensitivity of the 4FGL. The sample of detectable sources is used as verification: both the cumulative spectrum and the $\log N - \log S$ distribution of detectable sources are well matched in simulations and data in the energy range considered in the Fermi analysis. As model for the unresolved emission we use the mean of 200 realisations, which results in a flux in unresolved sources of around $1\%$ of the detected sources. Individual realisations exhibit large fluctuations, manifested in a standard deviation in the simulated fluxes of $>100\%$.  
    \item {\it the secondary emission of Galactic CRs}: at the energies relevant for this paper ($E \gg 10$ GeV) this can be further divided in the hadronic component, due the decay of neutral pion produced by CR scattering onto the IS gas (mostly hydrogen and Helium), and the Inverse Compton (IC) emission of CR electrons and positrons onto the ISRF.    
\end{itemize}
The relative contributions of these components depend on the Galactic coordinates and on the energy. On the GP and at energies larger than 10 TeV, the hadronic emission by CRs is expected to be dominant although a significant -- see \citet{Linden:2017blp}  -- contribution due to IC emissions cannot be excluded. 

Here we focus mainly on modeling the secondary diffuse emission due to interaction of Galactic CRs during their propagation. 
We do that with the {\tt HERMES} \citep{Dundovic:2021ryb} code which, at each given energy bin and for each relevant CR species, performs a numerical integration along the line-of-sight of the product of the CR differential energy flux, of the IS gas density (or the ISRF for the IC emission) and of the $\gamma$-ray production cross section.  
More details on the cross-sections and the gas (Hydrogen and Helium) distributions used in this work will be given in Secs.~\ref{sec:CRtransp} and~\ref{sec:gas} respectively. 

In the following subsection we rather discuss how the CR energy and spatial distributions are computed.  

\subsection{The interstellar gas}\label{sec:gas}

Our model consists of a set of column density maps in $(l,b)$ Galactic coordinates for atomic and molecular gas, associated to Galactocentric rings.
The atomic gas model is based on the 21-cm line emission data observed by the recent HI4PI survey that covers the whole sky with a 1/12 degree binning \citep{HI4PI}. 
As far as molecular gas is concerned, the decomposition is based on the observations of the CO rotational line at $115$ GHz from the CfA survey \citep{Dame2001apj,Dame2004aspc}.
The profile decomposition is discussed in \citet{Galview1,GalDifModels}.
In our framework, every Galactocentric ring can be associated to a value of the CO-H$_2$ conversion factor ($X_{\rm CO}$). In our model, we adopt the values of [$1.8$, $3.5$, $4.0$, $4.5$, $7.5$, $8.0$] in units of $10^{20}$~cm$^{-2}$ K$^{-1}$ / (km s$^{-1}$) in the following Galactocentric radial intervals: [$0 - 3$ kpc; $3 - 5$ kpc; $5 - 6$ kpc; $6 - 7$ kpc; $7 - 15$ kpc; $15 - 30$ kpc].
We assume here that the ISM gas is a mixture of Hydrogen and Helium nuclei with uniform density ratio $f_{\rm He} = 0.1$.

\subsection{CR transport: the conventional and $\gamma$-optimized scenarios}\label{sec:CRtransp}

We determine the energy and spatial distribution of each relevant CR species solving numerically the transport equation with the {\tt DRAGON2} code \citep{Evoli2017jcap,Evoli2018jcap}.
We assume that the observed CR spectrum can be approximated as a steady-state solution for a smooth distribution of continuous sources, 
which we fix on the basis of SNR catalogues (here we use the SNR distribution reported in \citet{Ferriere:2001rg}).
For a given source spectrum -- generally a broken power-law tuned against locally measured CR spectra -- as an output the code provides the propagated spectra of each primary and secondary species in each point of the Galaxy. 
Besides several astrophysical quantities, as an input the code needs to receive the CR diffusion coefficient $D(\rho, {\vec x})$ as a function of the particle rigidity $\rho$ and of the spatial coordinates. In the conventional scenario this is assumed to be a single power law function of the particle rigidity with a spatially dependent slope, parameterized as follows:
$$
D(\rho, {\vec x}) = D_0 \cdot \beta \left(\frac{\rho}{\rho_0}\right)^{\delta({\vec x})},
$$
where $D_0$ is its normalization at a reference rigidity $\rho_0 =4\, \mathrm{GV}$\footnote{Often, for simplicity, $D_0$ is assumed to be spatially independent.}, and $\beta$ is the velocity of the particles in units of the speed of light. The index $\delta$, \textit{a priori} being poorly known, is inferred from the comparison with the measured secondary to primary CR flux ratios, the boron-to-carbon (B/C) ratio being the most common.
Works based on multi-channel analysis~\citep{Genolini:2019ewc, Fornieri2020jcap, LuqueMCMC} of AMS-02 results \citep{AMS:2016brs}, including others based on antiprotons data \citep{DiBernardo:2009ku, Luque:2021ddh}, found that at the Solar System $\delta(R_\odot) \simeq 0.5$.
A different scenario arises if $\delta = \delta({\vec x})$ which turns into a {\it non-factorized} dependence of the propagated CR spectra on energy and position. 
For the models studied here, the Alfv\`en velocity is taken to be $V_A = 13$~km s$^{-1}$, the normalization of the diffusion coefficient is $D_0 = 6.1 \times 10^{28}$~cm$^2$s$^{-1}$ and the halo size is $H=6.7$~kpc, in agreement with recent analyses of $^{10}$Be ratios \citep{DeLaTorreLuque:2021yfq}. We checked that passing to the $\gamma$-optimized scenario has no effect on the local B/C (see e.g. \citet{Gaggero:2014xla}), as well as on other secondary-to-primary CR ratios, which indeed are correctly reproduced with this setup.
We notice that adopting a different value of $H$ and rescaling $D_0$ so to keep their ratio -- hence the CR grammage -- unchanged would have no significant effect on the CR propagated spectra and the $\gamma$-ray diffuse emission along the GP.  Concerning the vertical -- with respect to the GP -- $z$ dependence of the diffusion coefficient, we assume here a simple step function $D(\rho,R,z) = D(\rho,R) \Theta({ H - \vert z \vert})$. Since the target gas is concentrated in a thin disk with $z_{\rm gas} \ll H$, varying the value of $H$ within errors, or using a smoother vertical profile, would not have significant effects on the line-of-sight integrated $\gamma$-ray emission. 

Probing only the spectra of CR reaching the Solar System, the detection of charged CRs does not allow to discriminate among the factorized (conventional) and non-factorized scenarios. 
Indeed, the first evidence supporting the latter scenario was found in \citet{Gaggero:2014xla} on the basis of the Fermi-LAT results showing an excess of the diffuse $\gamma$-ray emission of the Galaxy above $10$~GeV in the inner GP respect to the predictions of the conventional one. 
In that work it was shown that the Fermi-LAT results are reproduced if $\delta$ has a linear dependence on the Galactocentric radius $R$ which turns into a harder spectrum of CR protons at low $R$, hence of secondary $\gamma$-rays at low Galactic longitudes (see also \citealt{Fermi-LAT:2016zaq}).
For this reason, in the following we will call this ${\bf \gamma}$-{\it optimized} scenario.
Afterwords, it was shown \citep{Cerri2017jcap} that this scenario is theoretically motivated and arises as a consequence of the growing poloidal component of the Galactic magnetic field at small Galactocentric radii (see \textit{e.g.} \citealt{Jansson2012apj}).

In the following we will consider two transport setups: the {\it Base} one, which is representative of the conventional scenario, and the $\gamma$-{\rm optimized} one. The main parameters of those models are reported in Table \ref{tab:diff_params}. 
For the $\gamma$-optimized setup we find

$$
\delta(R) = 0.04({\rm kpc^{-1}}) \cdot R({\rm kpc}) + 0.17,
$$

\noindent for $R < R_\odot = 8.5$ kpc and 
$\delta(R) = \delta(R_\odot)$ for $R \geq R_\odot$.
% 0.04({\rm kpc^{-1}}) R({\rm kpc}) + 0.033
As discussed at the end of Sec.~\ref{sec:CRdata}, for each model we will use two different choices -- Min and Max  setups -- of the CR proton and Helium source spectra, which enter to determine the spectra propagated to the Earth, so to bracket the experimental uncertainties above 10 TeV (see Fig.~\ref{Fig1}).
{\color{black} This arrangement effectively accounts also for the scenario in which the features in the propagated spectra are originated by transport rather than by the acceleration mechanism close to the sources.}
%While the Min models are built to match the KASCADE and KASCADE Grande data (which it does very well both for protons and Helium) and it also better traces DAMPE at lower energies, the MAX models are better suited to reproduce IceCube protons (although it overshoot Helium data above the PeV that has a minor impact on the $\gamma$-ray spectrum).  
CR electron source spectra are also tuned to match local experimental data up to the few TeV's -- see Fig. \ref{fig:ElectronSpectrum} in the Appendix \ref{sec:ElAppendix}.

\begin{table*}[t]
    \centering
    \begin{tabular}{|c|c|c|c|c|c|c|c|c|}
    \hline
    \multicolumn{9}{c}{\textbf{Injection parameters}} \\
    \hline 
    \hline
    & $^1\mathbf{H} \; \gamma_1$ & $^1\mathbf{H} \; \gamma_2$ & $^1\mathbf{H} \; \gamma_3$ & $^1\mathbf{H} \; \gamma_4$ & $^4\mathbf{He} \; \gamma_1$ & $^4\mathbf{He} \; \gamma_2$ & $^4\mathbf{He} \; \gamma_3$ & $^4\mathbf{He} \; \gamma_4$ \\
    \hline
    Max model & 2.33 & 2.23 & 2.78 & --- & 3.28 & 2.18 & 2.69 & --- \\
    \hline
    Min model & 2.33 & 2.16 & 2.44 & 3.37 & 2.30 & 2.06 & 2.34 & 3.01 \\
    \hline
  \end{tabular}
  
  \caption{\small{Spectral indexes at injection for the Max and Min models. These spectral indexes are tuned to CR local data as described above and correspond to spectral breaks at the following energies: $335$ and $6 \cdot 10^6$~GeV for the Max models and $335$, $2\cdot 10^4$ and $4 \cdot 10^6$~GeV for the Min models. }}
    \label{tab:diff_params}
\end{table*}

Both models are implemented with the {\tt DRAGON2} code \citep{Evoli2017jcap,Evoli2018jcap} -- which was designed to enforce spatially dependent diffusion --  using the same cross-sections and IS gas and ISRF as discussed below.
Hadronic emission maps due to CR interactions on IS Hydrogen and Helium, as well the IC emission by electrons, are then computed with the {\tt HERMES} code~\citep{Dundovic:2021ryb}. We adopt the $\gamma$-ray production cross-section described in~\citet{Kelner2008prd} with the updated parameterization of the proton–proton total inelastic cross-section reported in~\citet{Kafexhiu2014prd}.
For Helium, we assume that it contributes with a geometrical weight estimated considering its atomic radius as a function of the atomic mass $A$, $R(A) \simeq A^{1/3} \, R_0$, being $R_0$ the reference proton radius. Since its contribution appears due to the $pp$ interaction cross-section, then its geometrical weight becomes $R(A_{\mathrm{He}}) \big/ R(A_p) \simeq \left( A^{1/3}_{\mathrm{He}} \right)^{2} = 4^{2/3}$, with respect to protons.
At high energies the absorption by pair-production on the spatially-independent CMB photon field is computed as shown in ~\citealt{Dundovic:2021ryb}. The effect of the infrared Galactic Background is negligible compared to the uncertainties and we neglect it in this work.

\begin{figure}[h!]
   \centering
   \includegraphics[width=\linewidth]{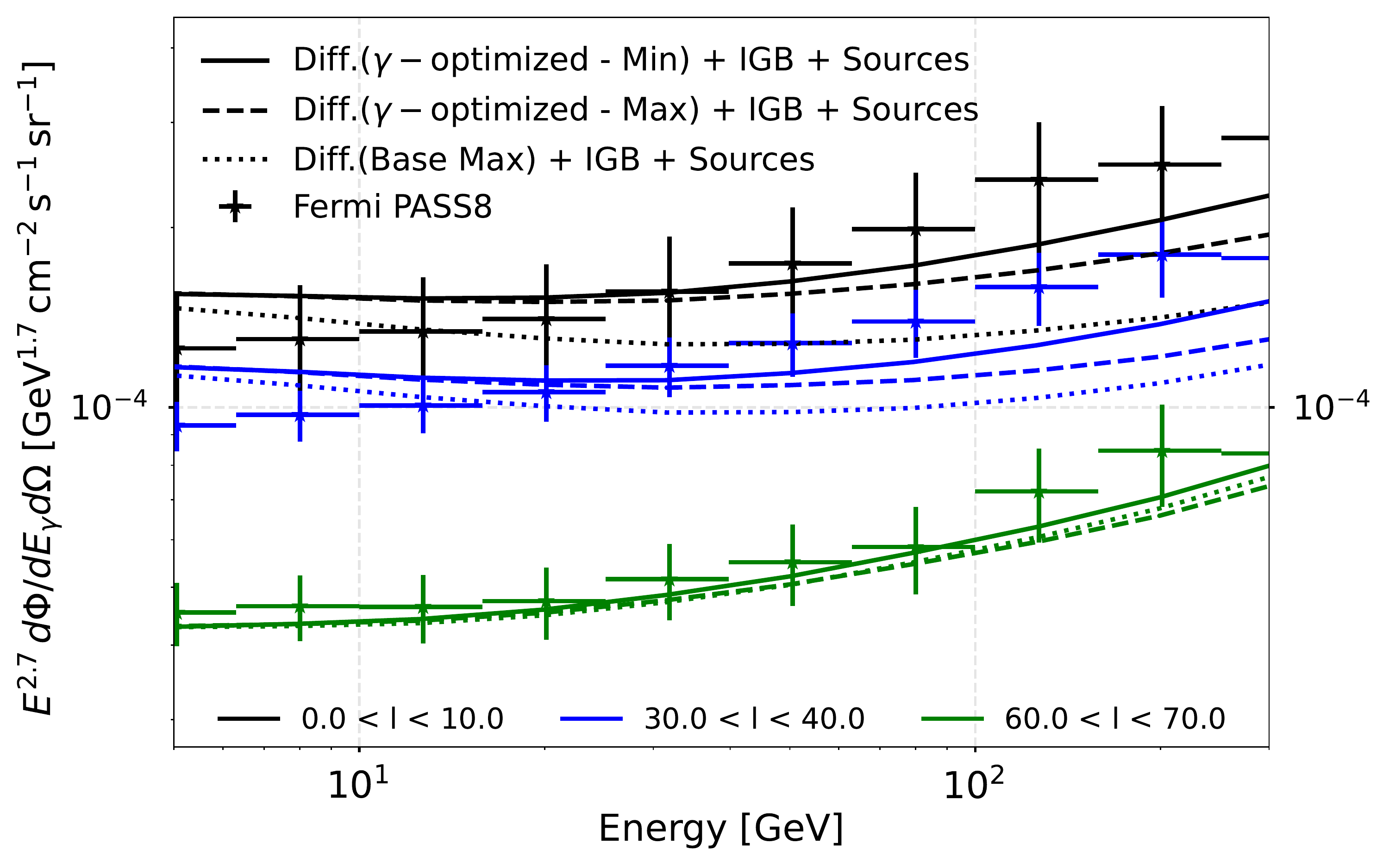}
   \caption{\small{The average spectrum of the $\gamma$-ray diffuse emission along the galactic plane ($| b | < 5^\circ $) is compared with Fermi-LAT data in three longitude intervals. The ``Sources'' component comprises the sources included in the 4FGL Fermi Catalog as well as unresolved sources and the Interstellar Galactic background light (``IGB'') component comprises the extra-galactic background light and Fermi's instrumental background. The errorbars represent just the statistical error of the measurements.}}
\label{Fig2}%
\end{figure}

In Fig.~\ref{Fig2} we compare our models with the spectrum of the diffuse emission below 300 GeV obtained from Fermi-LAT data as described in Sec.~\ref{sec:Fermi_data}. Not to overcrowd the figure, for the {\it Base} transport model only the Max setting is shown.  
A more comprehensive comparison is presented in Fig.~\ref{fig:FermiComparison} in Appendix \ref{sec:FermiAppendix}.
As mentioned in the above, the $\gamma$-optimized setups provide a better agreement with Fermi-LAT results a low Galactic longitudes.    

We notice that simplified versions of the Base and $\gamma$-optimized models considered in this work -- adopting an exponential cutoff at $5\,{\rm PeV/n}$ and neglecting absorption -- where used in \citet{CTA:2020qlo}.

%These models allow us to bracket the experimental uncertainties at PeV energies using two different interpretations of the spectra measured below this energy.

%\subsection{Gamma-ray production cross-sections and absorption}\label{sec:cs}
%{\textcolor{red} [Carmelo, Ottavio]}

%\subsection{Gamma-ray absorption in the Galaxy}\label{sec:absorption}

\vskip 1.cm 

\section{Results}

\begin{figure}[h!]
   \centering
  \includegraphics[width=\linewidth]{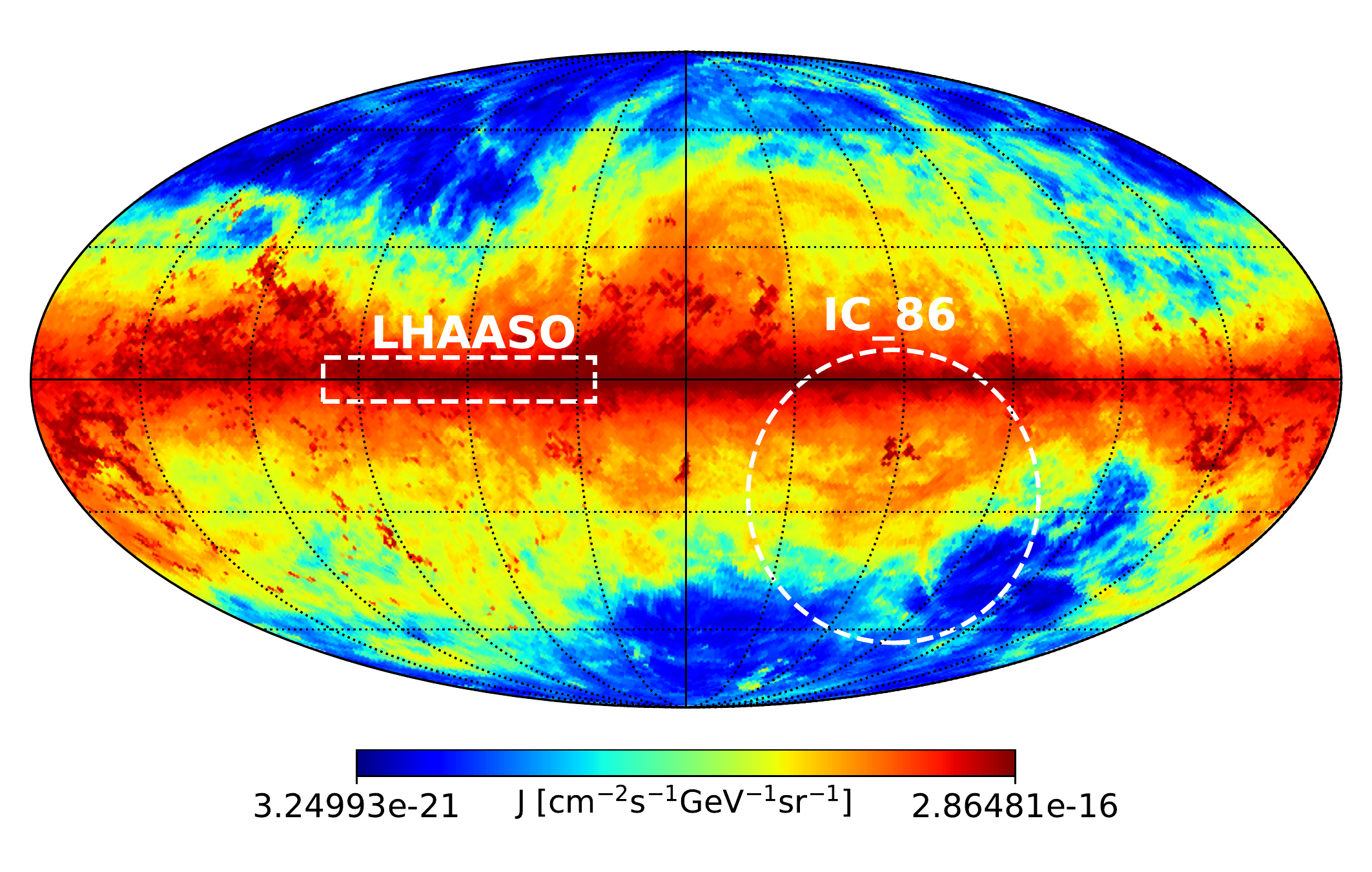}
   \caption{\small{The Mollweide projection of the all sky map of the Galactic diffuse emission flux above 100 TeV obtained for the Min $\gamma$-optimized model is reported in this figure. The contours of the regions probed by Tibet/LHAASO and by IceCube (IC-86) are reported. $J$, in the legend, means the differential flux of $\gamma$-rays per unit of solid angle.}}
    \label{fig:fullskymap}%
\end{figure}
%As we mentioned in the above, for a given CR transport model the {\tt HERMES} code allows to get full sky maps of the $\gamma$-ray (as well as other messengers not shown here, see \citet{Dundovic:2021ryb}) diffuse emission of the Galaxy from sub-GeV up to PeV energies. 
%The angular resolution is limited by the accuracy of the adopted IS gas distribution. For the gas model used in this paper (see Sec. \ref{sec:gas}) it can be as good as .....
%In Fig.(\ref{fig:fullskymap}) we show the Mollweide projection, performed with {\tt HealPix} \citep{HEALPIX2005}, of the map obtained for the $\gamma$-optimized Min model using a lower resolution (...).  {\color{red} We also display in the same figure} the contours of the regions observed by LHAASO (coincident with Tibet AS$\gamma$ and ARGO) and IceCube-86 \footnote{Kindly provided by Hershal Pandya}. 
%For any of those regions, we obtain the mean energy spectra for some of the models considered in this work which we represent in Fig.s (\ref{fig:Tibet_LHAASO}) and (\ref{fig:IceCube_CASAMIA}) respectively and compare with the experimental data. 
%The absorption due to $\gamma-\gamma$ scattering is accounted as described at the end of Sec.\ref{sec:CRtransp}. Its effect is shown in Fig.(\ref{fig:absorption}) for the $\gamma$-optimized scenario.

 We compute the full-sky maps of the diffuse gamma-ray emission associated to $\pi^0$ emission, Inverse Compton scattering and Bremsstrahlung with the {\tt HERMES} code \citep{Dundovic:2021ryb}. 
We choose an angular resolution characterized by the {\tt Healpix} resolution pararameter {\tt nside} = $512$, corresponding to a mean spacing between pixel of $\simeq 0.11^{\circ}$ \citep{HEALPIX2005}, nicely matching the angular resolution of the gas models adopted to compute the hadronic emission.
For illustrative purpose, we show the Mollweide projection of the total emission associated to the $\gamma$-optimized Min model in Fig. \ref{fig:fullskymap}, in a lower resolution.

In order to directly compare our models to the different experimental results described above, we consider several regions of interest, directly associated to the spectral data provided by the experiments focused on the very-high-energy domain. In particular, we show in the same Figure the contours of the regions observed by LHAASO (coincident with Tibet AS$\gamma$ and ARGO) and IceCube-86.

We obtain the integrated flux in these regions, which we compare to the experimental data without any further {\it ad-hoc} tuning and post-processing. We emphasize once again that all the details of the setup (in particular, the ring-by-ring normalization of the molecular gas density, and the CR transport setup) are set by the comparison with both local data on charged CRs and Fermi-LAT data in the GeV-TeV domain, as commented in more details in the Appendix.
The results are presented in Fig.s (\ref{fig:Tibet_LHAASO}) and (\ref{fig:IceCube_CASAMIA}). The absorption due to $\gamma-\gamma$ scattering is accounted as described at the end of Sec.\ref{sec:CRtransp}. Its effect is shown in Fig. \ref{fig:absorption} for the $\gamma$-optimized scenario.

Fig. \ref{fig:Tibet_LHAASO}, in particular, clearly represents the main result of this paper. 
%Noticeably this plot displays a rough consistency between $\gamma$-ray data from 10 GeV up to the PeV and all the considered models which assume the diffuse emission of the Galaxy to be originated by the CR sea of the Galaxy.
 This plot demonstrates that the diffuse emission models presented in this work --- obtained under the assumption that the emission is fully originated by the diffuse Galactic CR ``sea'' --- are able to capture the main features of the observed data in a remarkably large range of energies, from 10 GeV all the way up to the PeV domain.
This is already a major result.

However, since we are willing to go beyond this first level of interpretation and use our results to learn something about Galactic CR properties we face two main problems:
\begin{itemize}
\item there is a significant degeneracy between the choice of the CR transport setup and that of the source spectra (which, as we shown, depends also on the CR data systematics);
\item there is a significant scatter of the Tibet and LHAASO data above 50 TeV. 
\end{itemize}

While this situation is likely to improve with the next data releases we may already get some valuable hints limiting ourselves to consider only the lowest energy bin of both experiments which should be affected by lower systematics. 
Interestingly we notice that the four lowest energy LHAASO points -- below 50 TeV -- are well aligned among themselves and the Tibet ones.
We notice that those data favour the $\gamma$-optimized Max model.  
Even if we were to disregard Tibet data, or assume them to be contaminated by the emission of the Cygnus cocoon (see Sec.~\ref{sec:gamma_data}), the $\gamma$-optimized scenario would remain the preferred one though in its Min realization (see also Fig. \ref{fig:absorption}).  Although the Base - Max model is also in reasonable agreement with LHAASO data it is disfavored by Fermi-LAT and ARGO results.  
This shows the importance of using data over the widest possible energy range.

\begin{figure}[h!]
   \centering
  \includegraphics[width=\linewidth]{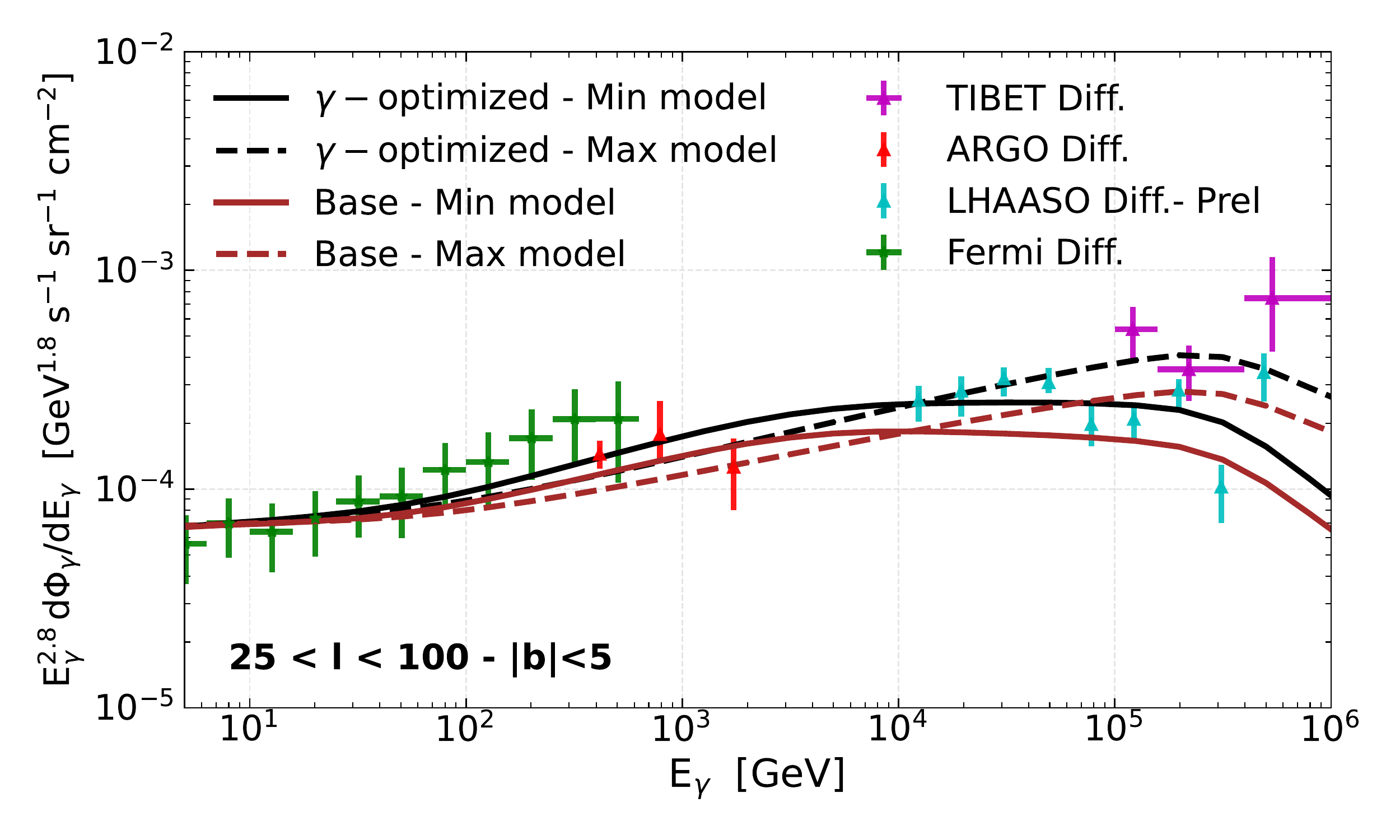}
   \caption{\small{The $\gamma$-ray spectra computed within the conventional (base) and $\gamma$-optimized scenarios are compared to Tibet AS$\gamma$ 
   \citep{TibetASgamma:2021tpz} and LHAASO \citep{Zhao:2021dqj} (preliminary) data in the window $| b | < 5^\circ $, $ 25^\circ < l < 100^\circ $. The Galactic diffusion emission spectrum measured by Fermi-LAT and extracted as discussed in Sec.~\ref{sec:Fermi_data}, as well as ARGO-YBJ data \citep{ARGO-YBJ:2015cpa} in the same region, are also reported.
%   The shadowed region are delimited by the Min e Max set-ups for each transport scenario.
   The models account for the effect of $\gamma$-ray absorption onto the CMB photons (see Sec.~\ref{sec:CRtransp}).}}
\label{fig:Tibet_LHAASO}%
\end{figure}

\begin{figure}[h!]
\centering
\includegraphics[width=\linewidth]{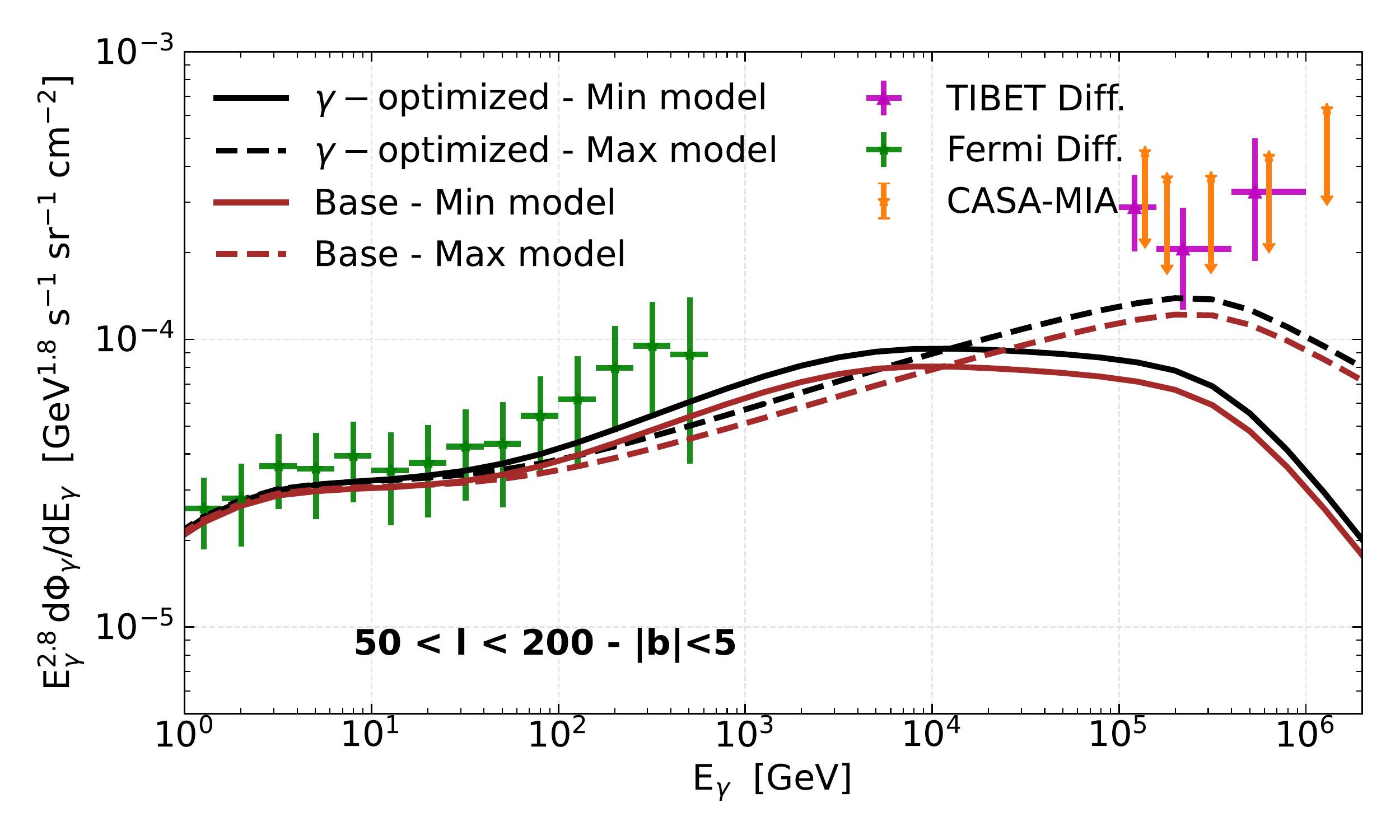}
\caption{Predicted $\gamma$-ray spectra for the different scenarios studied in this work and compared to Tibet AS$\gamma$ \citep{TibetASgamma:2021tpz} and Fermi-LAT data in the window $| b | < 5^\circ $, $ 50^\circ < l < 200^\circ $. The experimental errorbars show the $1\sigma$ statistical uncertainty of the measurement. Fermi-LAT systematic uncertainties dominate above $\sim200$~GeV, while the systematic error associated to TIBET data in this region is estimated to be around $30\%$~\citep{Amenomori2009}. 
CASA-MIA \citep{Borione:1997fy} upper limits in the same region are also reported.
%due to the uncertainty of absolute energy scale.
}
\label{fig:TIBET50-200}
\end{figure} 

We also consider the Tibet AS$\gamma$ data in the window $| b | < 5^\circ $, $ 50^\circ < l < 200^\circ $ (Fig. \ref{fig:TIBET50-200}). We notice that in this more external region the predictions of the $\gamma$-optimized and Base scenarios are quite similar so that those data may help to remove the degeneracy between the choice of the transport scenario and the shape of the source spectrum.  
Remarkably, even accounting for a possible contamination due to Cygnus-OB2, 
Tibet results seems to neatly favour the Max setup for the latter unknown.  
It will be very interesting, therefore, to see if LHAASO will possibly confirm Tibet results in that region. This will be also relevant to scrutinize an alternative interpretation of Tibet results given in terms of the emission of unresolved pulsar wind nebulae \citep{Vecchiotti:2021yzk}.

%In particular the $\gamma$-optimized Max model -- which interprets the spectral bump above 10 TeV as a local feature and connect CREAM III to KASCADE (2005 - QGSJET) and IceTop proton data (see Fig.(\ref{Fig1}))   
%-- which assumes a source spectral shape which allows to match the DAMPE bump and the proton and Helium KASCADE (Sybill MC) and KASCADE-Grande data %-- seems to be the one that gives the best agreement with those data.
%Above 60 TeV LHAASO preliminary data are more scattered while the highest energy Tibet's bin is likely to overestimate the diffuse emission as discussed in Sec.\ref{sec:gamma_data}.  

%In the first region we also report the experimental data including ARGO and Fermi-LAT the latter obtained as described in Sec.\ref{sec:Fermi_data}. For the latter region, since IceCube and CASAMIA refer to different regions, we rescaled them as described in \citet{IceCube:2019scr} (see their Fig.(12)).

We also performed a comparison of our models with IceTop and CASA-MIA upper limits which refer to regions different from those probed by Tibet and LHAASO (see Fig.\ref{fig:fullskymap}).
As evident from Fig. \ref{fig:IceCube_CASAMIA}, where we also report ARGO-YBJ data, although those limits do not constrain any of our models yet, the IceTop sensitivity is close to the level required to test the $\gamma$-optimized Max model. 

\begin{figure}[th!]
    \centering
    \includegraphics[width=\linewidth]{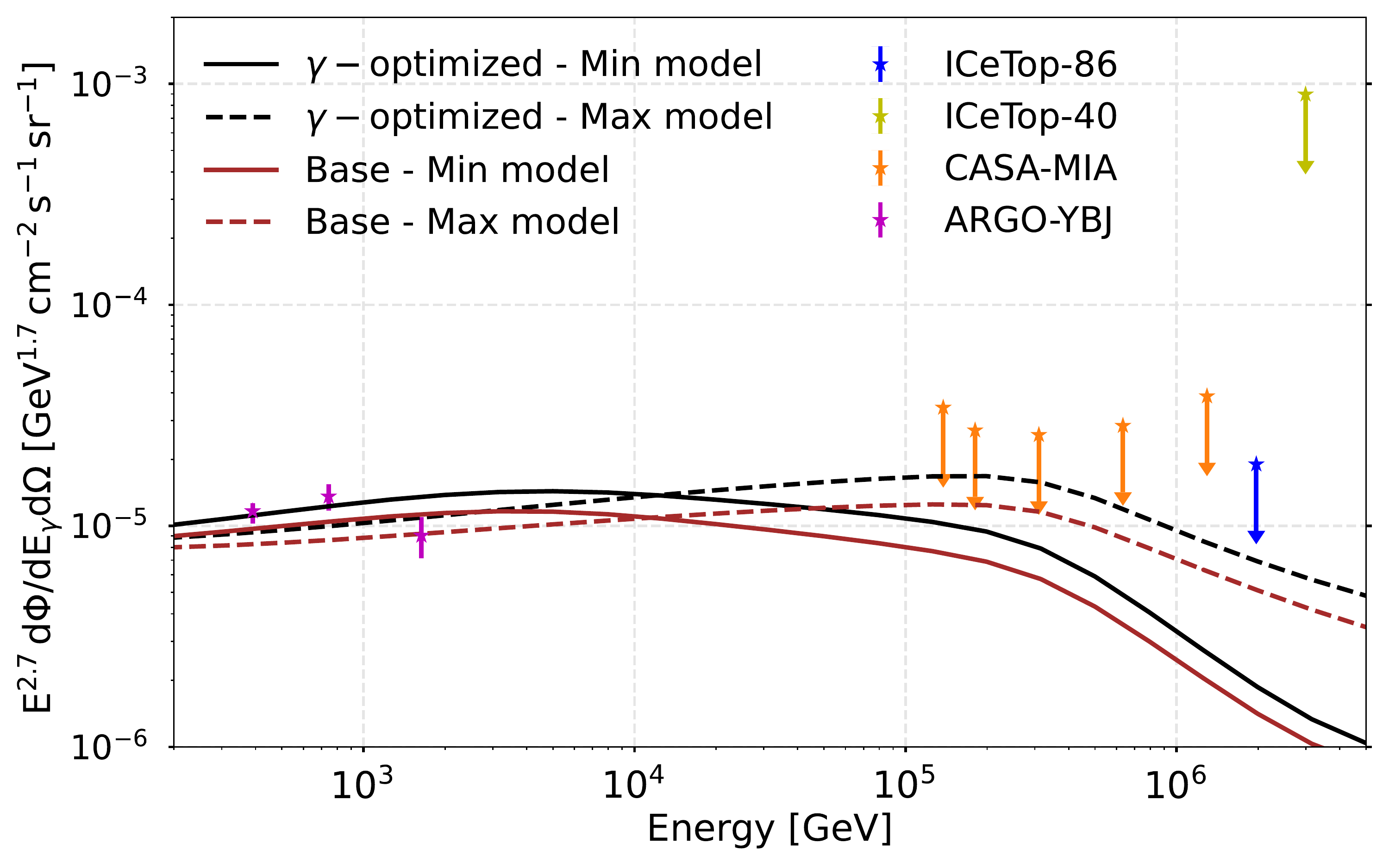}
    \caption{\small{The $\gamma$-ray spectra computed in the conventional (base) and $\gamma$-optimized CR transport scenarios are compared to IceCube \citep{IceCube:2019scr} and CASA-MIA \citep{Borione:1997fy} upper limits. Since those data refer to different regions of the sky, they are rescaled as described in \citet{IceCube:2019scr} (see Fig. 12 in that paper).}}
    \label{fig:IceCube_CASAMIA}%
\end{figure}

\begin{figure}[h!]
    \centering
    \includegraphics[width=\linewidth]{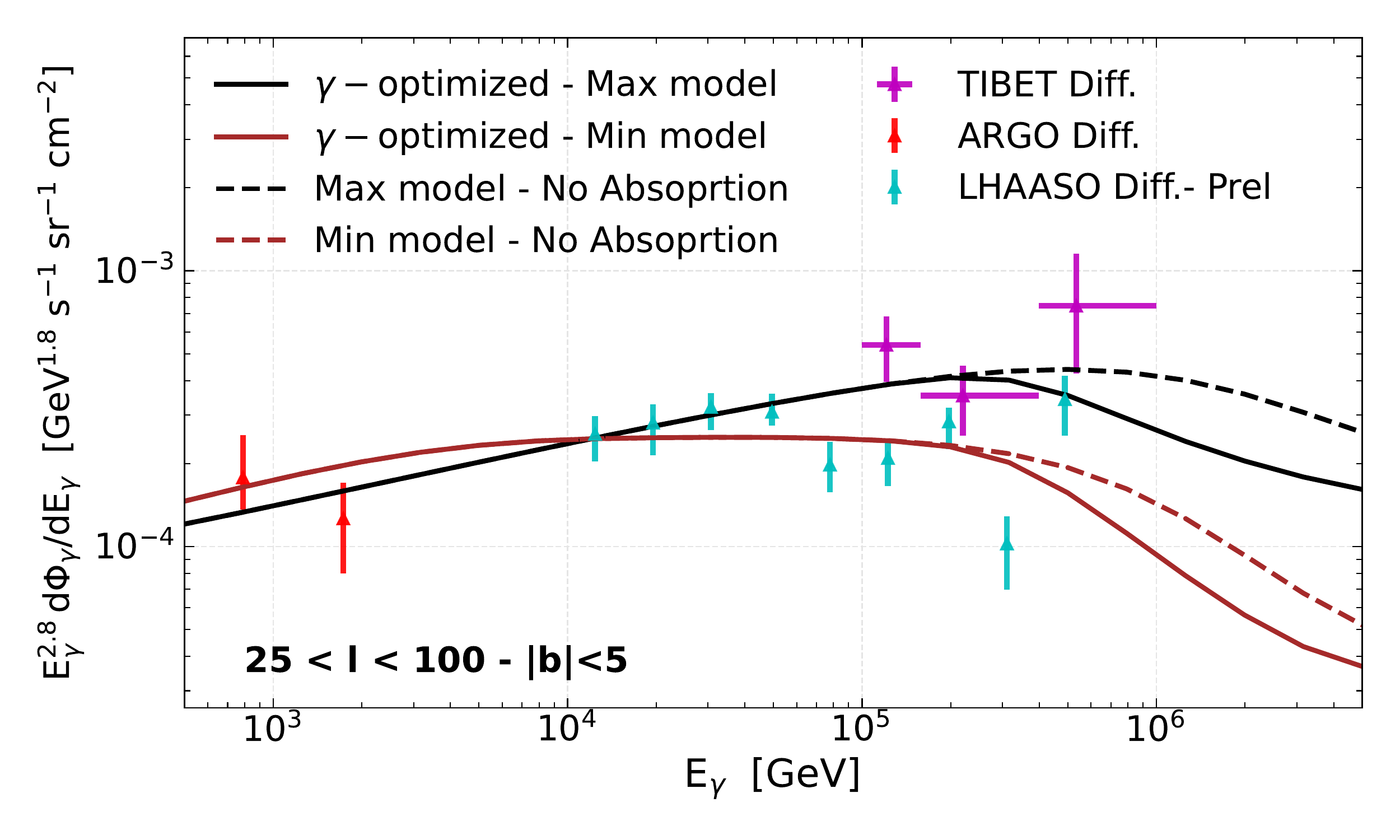}
    \caption{\small{In this figure we show the effect of $\gamma$-ray absorption onto the CMB photons (see Sec.~\ref{sec:CRtransp}) for the $\gamma$-optimized scenario.}}
    \label{fig:absorption}%
\end{figure}

\section{Conclusions}

In this paper we presented a set of gamma-ray diffuse emission models that are able to consistently reproduce the available measurements from the GeV domain all the way up to PeV energies in the Galactic Plane region.

In particular, we discussed a reference model based on the assumption of homogeneous transport properties in the whole Galaxy, and an optimized model aimed at capturing the progressive hardening of the proton slope inferred from Fermi-LAT data in the GeV domain. Both scenarios are tuned on local CR data, and are presented in two different versions, that correspond to a different fitting strategy of local CR data in the very-high-energy part of the spectrum, which results in different choices of the injection spectra.

We found a relevant degeneracy between the choice of the CR propagation setup and that of the injection spectrum. In spite of that, we argued that the comparison between our models and the combination of different $\gamma$-ray data sets is able to provide valuable hints and may help to shed light on CR transport properties in different regions of the Galaxy.

We highlighted in particular that the Galactic diffuse emission measured by most experiments from 10 GeV to the PeV can almost entirely be explained as truly diffuse emission stemming from the Galactic CR ``sea'', and showed that the Tibet AS$\gamma$ and LHAASO data, in combination with Fermi-LAT results, seem to favour a transport scenario characterized by spatially dependent diffusion.

%In this paper we discussed the crucial role that recently released and forthcoming $\gamma$-ray data may play in the understanding of  CR transport in the more distant regions of the Milky Way as well as the shape of CR protons and Helium spectra at energies as large as the PeV.  

%We argued that one of the major obstacles on that road is given by the degeneracy existing between the choice of the CR propagation model and that of the CR source spectra given also the large experimental uncertainties above 10 TeV. 
%In spite of those difficulties, we showed that combining $\gamma$-ray data of different experiments taken in different regions of the Galactic Plane, several valuable hints can be already obtained.

%Besides confirming that the Galactic diffuse emission measured by most experiments from 10 GeV to the PeV can -- within present uncertainties -- be explained by the emission of the Galactic CR sea, we showed that Tibet AS$\gamma$ and LHAASO in combination with Fermi-LAT results favour a transport scenario characterized by spatially dependent diffusion.

Our results give a stronger support to a similar claim of the Tibet collaboration -- based on the comparison of their data  with the phenomenological models in \citet{Lipari:2018gzn} -- which however did not take into account the degeneracy mentioned in the above. 

The comparison of our models with forthcoming experimental results at lower energies by HAWC \citep{Nayerhoda:2020ces}, H.E.S.S. \citep{2014PhRvD..90l2007A}, SWGO \citep{Albert:2019afb}, CTA \citep{CTA2019,CTAConsortium:2017dvg} and ALPACA \citep{Takita_2017,Sako:2021fyf} will be crucial to further validate this scenario and to probe the CR shape throughout the Galaxy, especially by those experiments, which will have a better view of its most central regions. 

In order to facilitate the comparison with these forthcoming data, we provide the scientific community with high resolution all-sky-maps of the diffuse $\gamma$-ray emission of the Galaxy from few GeVs to few PeVs computed for our benchmark models\footnote{\url{https://github.com/tospines/Gamma-variable_High-resolution}}. They can be valuable tools for experimental collaborations and can be used as ``background models'' in different contexts, from the generation of Galactic and extra-Galactic source catalogues to indirect dark matter searches. %in order to define a background for Galactic and extra-Galactic source catalogues production, for indirect dark matter search, besides as diffuse emission templates. 

As a final discussion point, let us return to the potential role of unresolved sources.
In general, the relative weight of {\it truly diffuse emission} and {\it unresolved source contribution} depends on a wide range of parameters, that characterize: the nature of the sources, the capability of the experiment to identify and resolve individual sources, the transport/escape of the high-energy particles that constitute the diffuse CR sea, and of course on the total amount of target gas and photon background that is directly responsible for the truly diffuse signal. 
In this work, we choose to consider a specific model for the unresolved contribution and take it into account consistently when tuning our prediction for the diffuse emission on the Fermi-LAT data. Due to the uncertainty associated in the modeling of this component, and the impossibility to directly extrapolate the model to the PeV domain (a task that would require an accurate characterization of each PeV-oriented experiment to resolve individual sources), we cannot exclude a degeneracy between our scenario and an alternative model featuring a more prominent role of the unresolved component.
%\melo{The relatively good agreement between our models -- where unresolved sources makes a fraction around 1\% of the total flux -- and experimental data suggest their role to be subdominant.}
However, in order to safely disentangle different interpretations,
%The disentanglement of these different interpretations 
%of the very-high-energy tail of the gamma-ray spectrum 
%clearly requires 
to have more data will be crucial. In particular, we emphasize the role of neutrino measurements in this context (see also  \citealt{Ahlers:2015moa,Gaggero:2015xza,Fang:2021ylv,Breuhaus:2022epo}). In fact, the truly diffuse component is likely to be entirely hadronic above few TeV, while the sources that could potentially contribute to the unresolved flux are mostly leptonic (namely, pulsar wind nebulae). Hence, a future firm discovery of a neutrino diffuse component associated to the Galactic plane would be a clear signature of the scenario we propose here.

\begin{acknowledgements}
We thank Quentin Remy for providing us with the interstellar HI and H$_2$ 3D distributions (``ring model'') used in this work. We also thank Hershal Pandya for informing us about IceCube collaboration $\gamma$-ray measurements and providing us with the corresponding sky window. 
We thanks Paolo Lipari and Silvia Vernetto as well, for reading our manuscript and giving us useful comments.
D.Gaggero\ has received financial support in the early stage of the project through the Postdoctoral Junior Leader Fellowship Programme from la Caixa Banking Foundation (grant n.~LCF/BQ/LI18/11630014) during the early stage of the project. D.Gaggero\ was also supported by the Spanish Agencia Estatal de Investigaci\'{o}n through the grants PGC2018-095161-B-I00, IFT Centro de Excelencia Severo Ochoa SEV-2016-0597, and Red Consolider MultiDark FPA2017-90566-REDC during the early stage of the project.
D.Gaggero also acknowledges funding from the “Department of Excellence” grant awarded by the Italian Ministry of Education, University and Research (MIUR) for the period October-December 2021.
D.Gaggero also acknowledges support from the INFN grant “LINDARK,” and the project “Theoretical Astroparticle Physics (TAsP)” funded by the INFN for the period October-December 2021. D. Grasso is also supported by TAsP.
D. Gaggero also acknowledges the support from Generalitat Valenciana through the plan GenT program (CIDEGENT/2021/017) staring from 1/01/2022.
C.E.~acknowledges the European Commission for support under the H2020-MSCA-IF-2016 action, Grant No.~751311 GRAPES 8211 Galactic cosmic RAy Propagation: An Extensive Study.
P. De la Torre is supported by the Swedish National Space Agency under contract 117/19. Infrastructure for Computing (SNIC) under project Nos. 2021/3-42 and 2021/6-326 partially funded by the Swedish Research Council through grant no. 2018-05973.
\end{acknowledgements}


%\clearpage
\bibliographystyle{aa}     
\bibliography{2021-hermes,software} 

\onecolumn
\begin{appendix}
\section{Detailed comparison to Fermi-LAT data.}
\label{sec:FermiAppendix}
We present here a detailed comparison between the gamma-ray diffuse emission models presented in this work and the Fermi-LAT data, with particular focus on the Galactic plane. We emphasize that our modeling aims at capturing at first order the large-scale spectral trend inferred from the data and highlighted in several papers~\cite{Gaggero:2014xla,Pothast2018jcap,Acero2016apjs}. We do not perform any post-processing or fitting of the templates, neither do we include specific models for extended emission from, for instance, star-forming regions.

We consider different regions of interest along the Galactic Plane, and compare the Fermi-LAT {\tt PASS8} data to the sum of the following components, for each model setup considered here.

\begin{itemize}
    \item A model for the $\pi^0$ emission obtained with {\tt HERMES} as described in the text. The rescaling factor $X_{\rm CO}$ is tuned on a ring-by-ring basis in order to match the data normalization in the $[1 - 10]$ GeV interval.
    \item A model for the leptonic Inverse Compton scattering onto the diffuse low-energy photon component, obtained with {\tt HERMES}.
    \item A model for the emission due to Bremsstrahlung, obtained with {\tt HERMES}.
    \item The total flux associated to the resolved point sources, taken from the fourth Fermi Large Area Telescope catalog (4FGL) \citep{Fermi4FGL}.
    \item A model for the unresolved flux contribution, described in Sec.~\ref{sec:models}. 
\end{itemize}

\begin{figure*}[bh]
\begin{multicols}{2}
    \includegraphics[width=0.48\textwidth]{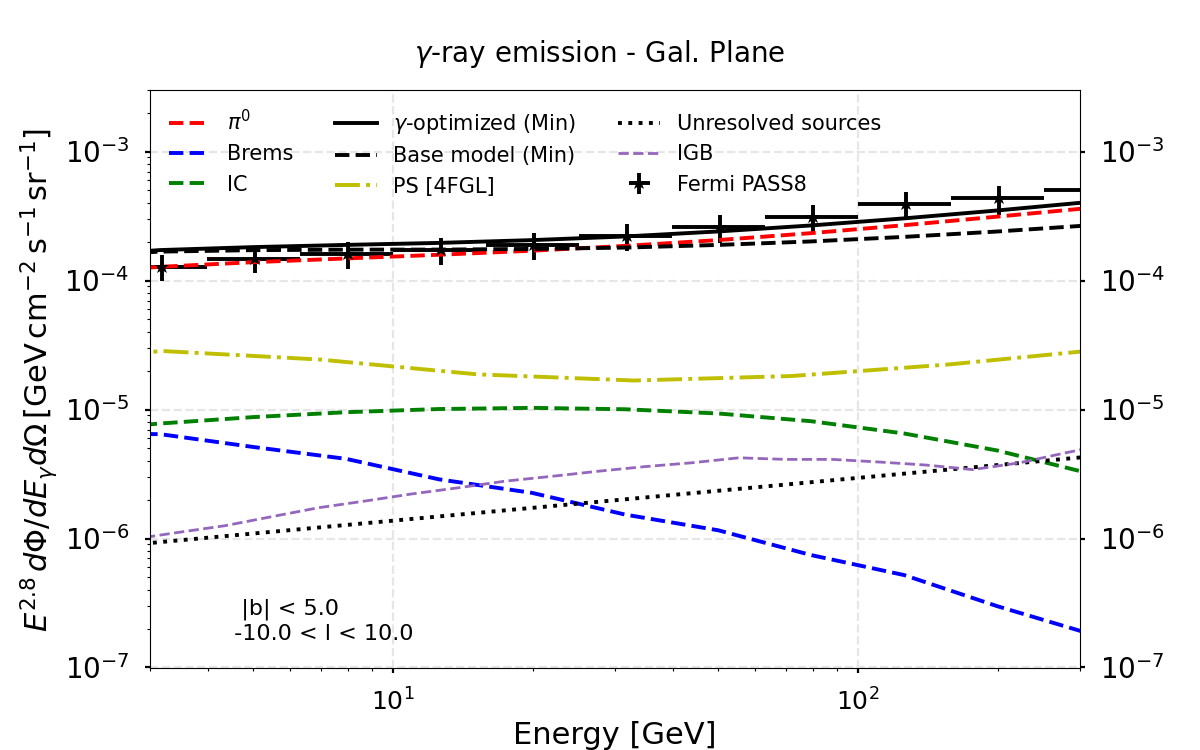}\par 
    \includegraphics[width=0.48\textwidth]{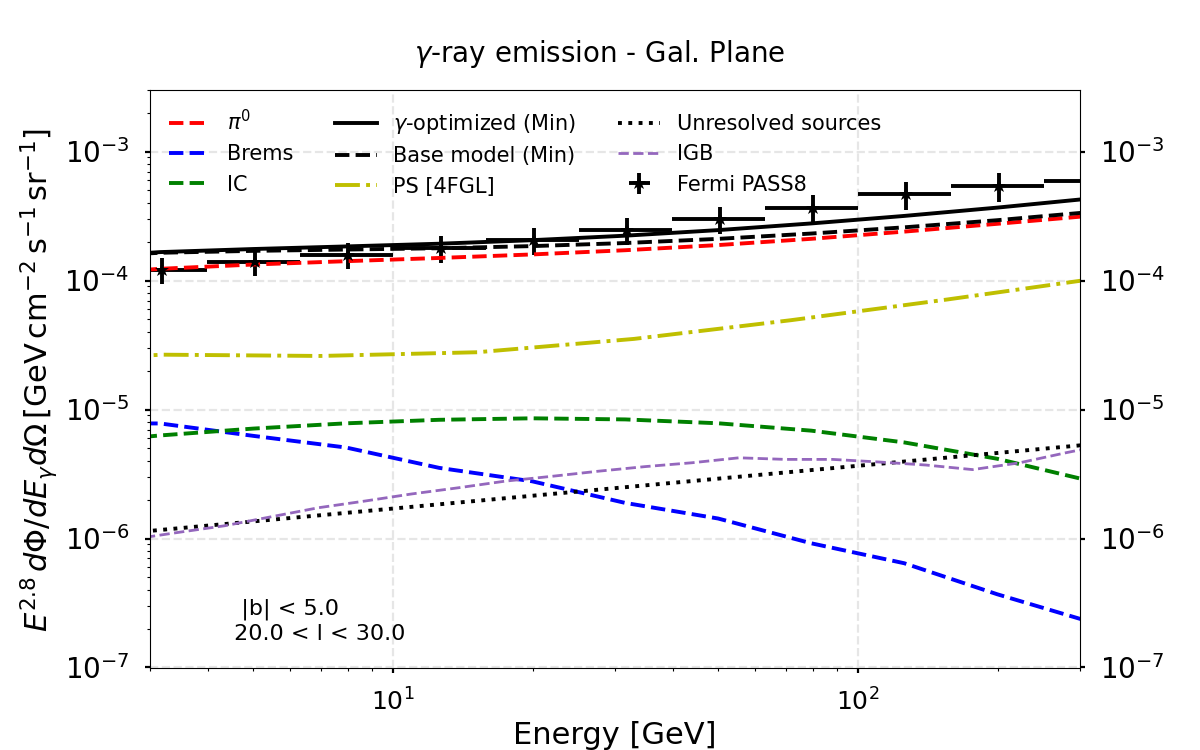}\par 
    \end{multicols}
\begin{multicols}{2}
    \includegraphics[width=0.48\textwidth]{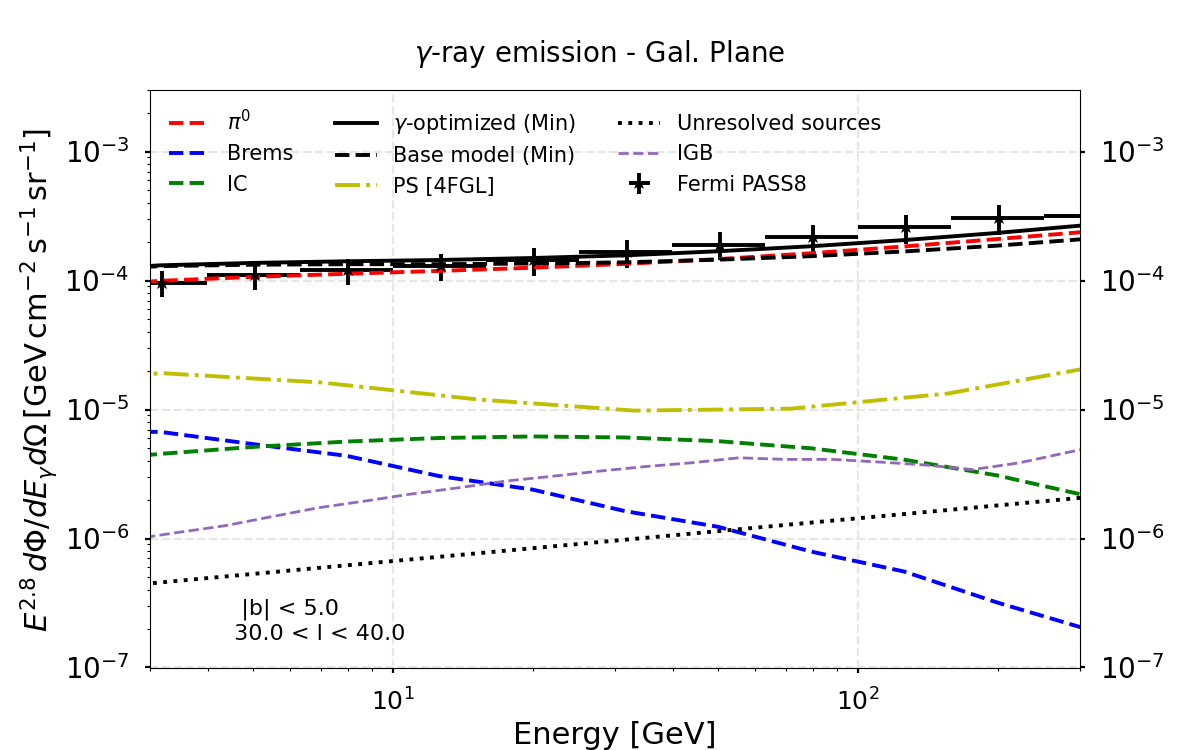}\par
    \includegraphics[width=0.48\textwidth]{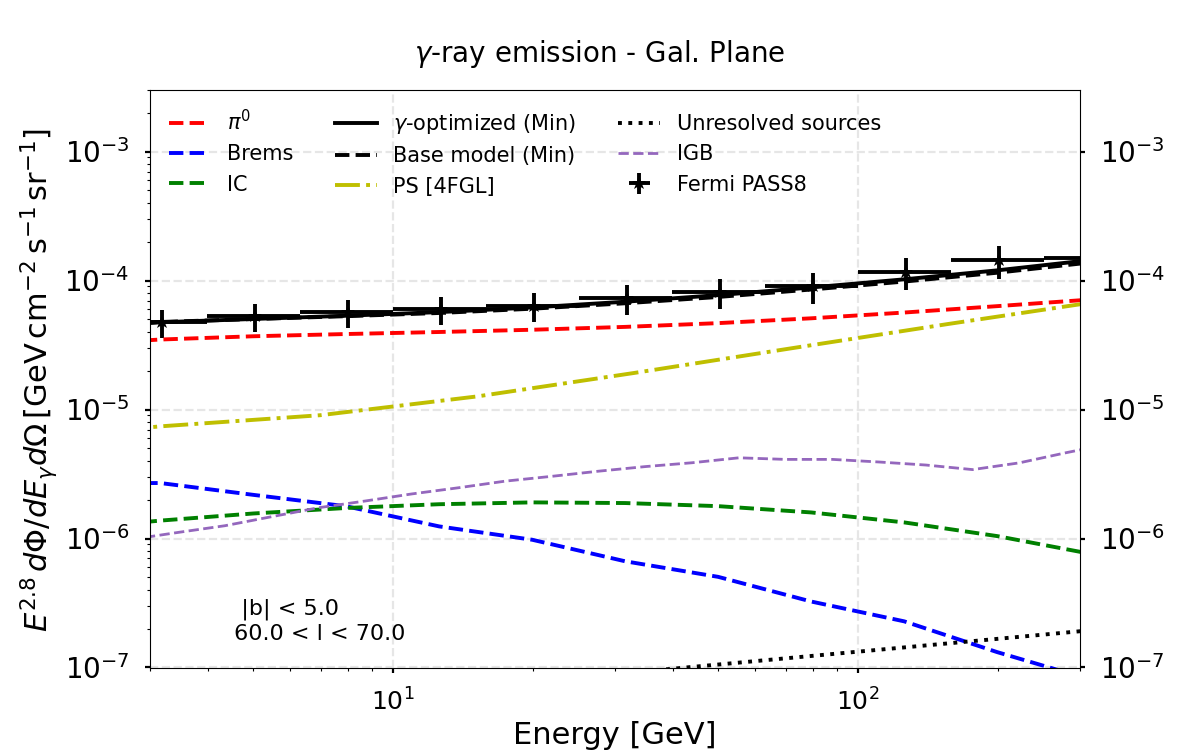}\par
\end{multicols}
\caption{Comparison between Gamma-optimized models and Fermi-LAT data.}
\label{fig:FermiComparison}
\end{figure*}

We show the comparison in Fig. \ref{fig:FermiComparison}. The three components of the Galactic Diffuse Emission (GDE) modeled with {\tt HERMES} are computed in both the {\it Base} (MIN) and the {\it Gamma-Optimized} (MIN) setups. In both cases, the total GDE is added to the isotropic spectral template and to the (unresolved+resolved) source model: The corresponding total contribution is plotted as a dashed (solid) black line for the Base (optimized) model.
We  notice an overall good accord of the gamma-optimized model with the data. In particular, the region characterized by $-10^{\circ} < l < 10^{\circ}$ is better reproduced by this setup, due to the hard spectrum detected by Fermi-LAT, confirming previous findings.

\newpage

\section{Electron spectrum}
\label{sec:ElAppendix}

\begin{figure*}[hb]
\centering
\includegraphics[width=0.6\textwidth]{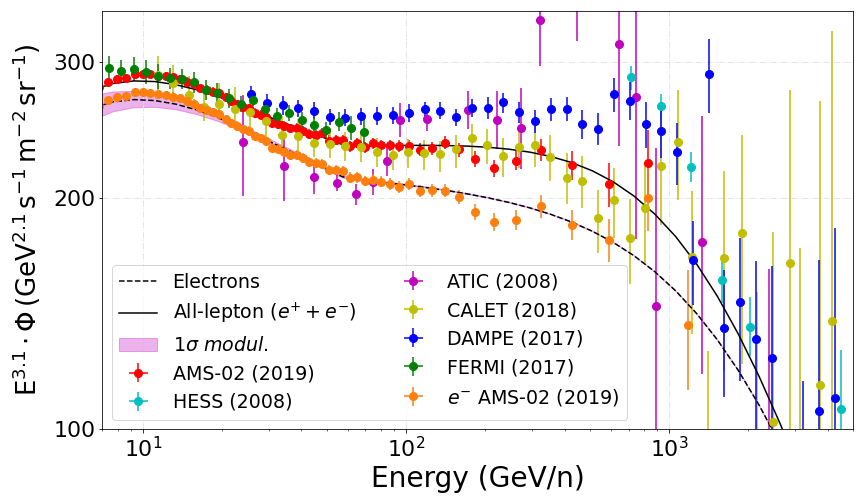}
\caption{Electrons and all-lepton spectrum for the models employed in this work, compared to available experimental data.}
\label{fig:ElectronSpectrum}
\end{figure*}

In this appendix, we show in Fig.~\ref{fig:ElectronSpectrum} the electrons and all-lepton spectrum computed with our models and fitted to experimental data from AMS-02, also in agreement with CALET~\citep{CALET_electrons}. We remark that the small discrepancy between AMS-02 data and the data measured by DAMPE does not affect our results, since this difference would mean a completely negligible contribution to the total diffuse emission.
A broken power-law with a break at $60$~GeV and spectral indices $\gamma_1=3.37$ and $\gamma_1=3.20$ (where $\gamma_1$, $\gamma_2$ are the spectral indices before and after the break, respectively) has been used to reproduce these spectra, with an exponential cut-off at $6$~TeV. 

\end{appendix}

\end{document}